\documentclass[%
 reprint,
 amsmath,amssymb,
 aps,
pra,
floatfix,
]{revtex4-1}
\usepackage{graphicx,lipsum}% Include figure files
\usepackage{dcolumn}% Align table columns on decimal point
\usepackage{bm}% bold math
\usepackage[dvipsnames]{xcolor}
\usepackage{array,multirow}
\usepackage{bm}% bold math
\usepackage{bbm}
\usepackage{amsmath}
\begin{document}

\preprint{APS/123-QED}
%\title{ Quantum coherence, non-Markovianity and speed limit time}
%\title{ The interplay between quantum coherence and mixedness for quantum speed limits for multi-qubit systems}
%\title{ Quantum coherence and mixedness of multi-qubit states for quantum speed limits}
%\title{ Quantum speed limit-a dynamical witness to distinguish  entangled multi-qubit states}
%\title{ Quantum speed limit time: Role of coherence  as a dynamical witness to distinguish multi-qubit entangled states }
\title{ Quantum speed limit time: role of coherence  }
%\title{Measure of non-Markovianity of non-unital and unital quantum dynamical maps}
%\title{Dynamics of quantum  correlations under the influence of  (non-)Markovian environmental interactions }% Force line breaks with \\

\author{K.G. Paulson\textsuperscript{a,b}}
\altaffiliation[paulsonkgeorg@gmail.com]{}%Lines break automatically or can be forced with \\
\author{Subhashish Banerjee\textsuperscript{b}}
\email{subhashish@iitj.ac.in }
\affiliation{Institute of Physics, Bhubaneswar-751005, India\textsuperscript{a}\\
Indian Institute of Technology Jodhpur, Jodhpur-342011, India\textsuperscript{b}}

%\textsuperscript{b,c}}
%\affiliation{
 %Third institution, the second for Charlie Author
%\affiliation{%


\date{\today}% It is always \today, today,

\begin{abstract}
The minimum evolution time between multi-qubit quantum states is estimated for non-Markovian  quantum channels. We consider the maximally coherent pure and mixed states as well as multi-qubit $X$ states as initial states and discuss the impact of initial coherence and the behaviour of coherence  on their speed of evolution for both dephasing and dissipative processes. The role of the non-zero value of initial coherence under information backflow conditions for the non-unital  dissipative process is revealed by the flow of quantum speed limit time ($\tau_{QSL}$). The trade-off between mixedness and coherence on the speed limit time reveals the nature of the quantum process the states undergo. The complementarity effect between mixedness and coherence is more prominent in the quantum non-unital dissipation process. The parametric trajectory of speed limit time vividly depicts the difference in the evolution of pure and mixed initial states, and this could be used to distinguish between the unital and non-unital channels studied in this work. Our investigation of  quantum speed limit time on multi-qubit  entangled $X$ states reveals that $\tau_{QSL}$ can be identified as a potential dynamical witness to distinguish multi-qubit states in the course of evolution. 

%\begin{description}
%\end{description}
\end{abstract}

\maketitle

%\tableofcontents

\section{\label{sec:level1} Introduction } 
Quantum speed limit time sets the lower bound for the evolution time between two quantum states. Different from the interpretation of the position-momentum uncertainty principle as the impossibility of simultaneous measurement of canonically conjugate variables, the energy-time uncertainty principle sets the time scale of quantum state evolution~\cite{mandelstam1945}. Therefore Heisenberg's energy-time uncertainty principle provides the bound on the minimum time required to evolve between two quantum states. Mandelstam, Tamm (MT) and  Margolus, Levitin  (ML)~\cite{margolus1998} derived bounds on the minimum time needed for a quantum system to evolve between the states $\tau_{QSL}$. The combined bound of  $\tau_{QSL}$ for an isolated system based on the square root of the variance and the expectation values of Hamiltonian is $\max\Bigg\{\frac{\pi\hbar}{2\Delta H},\frac{\pi\hbar}{2\langle H\rangle-E_{0}}\Bigg\}$, $E_{0}=0$ is generally considered as the ground state energy. The coupling of the system of interest with its surroundings reveals many intriguing characteristics of its dynamics. Accordingly, it's of interest to understand $\tau_{QSL}$ within the preview of open quantum systems. Different methods based on the measures of the closeness of quantum states like purity~\cite{del2013,uzdin2016speed}, fidelity ~\cite{deffner2013quantum, deffner2017} are adopted to derive MT-ML type bounds on speed limit time for open quantum systems. The quantum system dynamics in the non-Markovian regime is an active area of research for multiple reasons~\cite{jyrki1,rivas2014,de2017dynamics,li2018concepts,sbbook,utagi2020ping}. It's known that quantum correlations are lesser susceptible to noise in non-Markovian~\cite{paulson2021} realms, and quantum memory speeds up the evolution of quantum states~\cite{paulson2021effect}. These cases bring out  the need for detailed investigations of  quantum systems in non-Markovian environments. In this context, it would be of interest to note that there are a number of interesting works in the literature on the intersection of quantum speed limit time, non-Markovian physics and quantum technologies~\cite{Meng2015, Mukherjee_2015, dehdashti, Deffner_2017, Aggarwal_2022, zhen-yu, Campbell_2018, Dehdashti2020, riya-phase, Teittinen_2019}. 
\newline

Quantum coherence and mixedness play a vital role in the dynamics of quantum evolution. The presence of quantum coherence prompts the emergence of many intriguing properties of the quantum system. Quantum coherence and quantum correlations as a resource play a  pivotal role in a wide range  of fields; quantum information processing, quantum metrology, and quantum thermodynamics~\cite{suzuki2016,paulson2017,banerjee2017characterization,george2018,alicki2018introduction,bhattacharya2018evolution,liu2019quantum,paulson2019,dixit2019study}. In the literature, two different approaches have been proposed to quantify  coherence; viz. coherence as asymmetry relative to a group of translations, such as phase shifts or time translations~\cite{marvian2014modes,marvian2014extending}, and coherence as the resource relative to the set of incoherent operations~\cite{baumgratz2014}. In this work, we use the latter measure of coherence, and its interplay between mixedness is analyzed in the context of quantum speed limit time for single and multi-qubit states.\newline
% \textcolor{blue}{Characterization of quantum correlations of multipartite quantum systems from both the fundamental and application point of view is required for a number of quantum information tasks. An appreciable amount of research work  has already been devoted to the study of different measures of multipartite quantum correlations and their usefulness as successful resources to implement various quantum information and computation protocols (references). 
% We use the multi-qubit quantum resource to implement various quantum information and computation tasks efficiently} \newline

In the present work, we  investigate the role of quantum coherence and mixedness on  $\tau_{QSL}$ for pure, mixed as well as  multi-qubit $X$ states. We investigate $\tau_{QSL}$  based on relative purity and fidelity (Bures angle) for dephasing and dissipative  non-Markovian quantum channels. We show that for a pure dephasing channel, an increase in quantum coherence increases the $\tau_{qsl}$ between initial and final states. Our investigation on entangled $X$ states and maximally coherent entangled states reveals that the connection between the coherence and speed of evolution is not straightforward for the dissipation process.  The parametric trajectory of quantum speed limit time depicts the trade-off between coherence and purity as the evolution of the system progresses in various processes.  We also show the usefulness of quantum speed limit time as a witness for quantum state discrimination. Our analysis of $\tau_{QSL}$ on a class of maximally entangled multi-qubit $X$ states of equal initial coherence demonstrates that quantum speed limit time can be identified as a witness for quantum  states among different degenerate sets of entangled states. The action of a unitary operator transforms maximally entangled states into maximally coherent entangled states. We show that the quantum states cannot be distinguished upon maximizing the coherence. \newline
The present work is organized as follows. In Sec.~II, prerequisites for the current work are laid down, followed by a discussion on different measures of quantum speed limit time in Sec.~III. The role of coherence and mixedness on $\tau_{QSL}$ for various initial states under dephasing and dissipative processes are analyzed in 
Sec.~IV. Analytical calculations for coherence-dependent speed limit time for multi-qubit $X$ states are given in Sec.~V. We discuss the usefulness of $\tau_{QSL}$ as a potential witness to distinguish quantum states in Sec.~V, a theme which is followed up in Sec.~VI. In the end, we make our conclusion. The appendix contains the details of non-Markovian quantum channels used in this work. In particular, among the dephasing, unital class of channels, we consider modified Ornstein-Uhlenbeck noise (OUN) and random telegraph noise (RTN). In the dissipative, non-unital class, we discuss the non-Markovian amplitude damping channel (NMAD).

\section{Prelimenarires}
\subsection{Quantification of Coherence}
Here, we discuss the quantification of coherence based on the seminal work by Baumgratz et al. ~\cite{baumgratz2014}, in which the measure of coherence of state is defined as its deviation from incoherent states (a diagonal state in a fixed basis). A maximally coherent state is written as $\vert\psi\rangle=\frac{1}{\sqrt{d}}\sum_i^d\vert i\rangle$. The  key criteria every measure of quantum coherence $\mathcal{C}(\rho)$ is as follows: (I)
 $\mathcal{C}(\rho)\geq0$ for all quantum states and is zero for incoherent states $(\rho\in\mathcal{I})$.(II) Monotonicity under all the incoherent maps $\Lambda:\mathcal{C(\rho)}\geq\mathcal{C}(\Lambda(\rho))$, where $\Lambda(\rho)=\sum_{k}\mathcal{E}_{k}\rho\mathcal{E}_{k}^{\dag}$, $\{\mathcal{E}_{k}\}$ is a set of Kraus operators, $\sum_{k}\mathcal{E}_{k}\mathcal{E}_{k}^\dag=1$ with $\mathcal{E}_{k}\mathcal{I}\mathcal{E}_{k}^\dag\subset\mathcal{I}$. (III) Monotonicity for average coherence under subselection based on measurement outcomes: $\mathcal{C}(\rho)\geq\sum_{k}p_{k}\mathcal{C}(\rho_{k})$, where $\rho_{k}=\frac{\mathcal{E}_k\rho\mathcal{E}_k^{\dag}}{p_k}$, $p_k=Tr(\mathcal{E}_k\rho\mathcal{E}_k^{\dag})$ for all $\mathcal{E}_{k}$. (IV) Non increasing of coherence under convex combination of quantum states:$\sum_k p_{k}\mathcal{C}(\rho_{k})\geq\mathcal{C}(\sum_{k}p_{k}\rho_{k})$ for any combination of $p_{k}$ and $\rho_{k}$.  We have different  measures to quantify the coherence of quantum states that satisfy the above set of criteria (I-IV). Relative entropy is one among them, $\mathcal{C}_{re}(\rho)=S(\rho_{\textrm{diag}})-S(\rho)$, where $\rho_{\textrm{diag}}$ is the matrix containing only diagonal elements of $\rho$ in the reference basis and  $S(\rho)=-\textrm{tr}(\rho \log\rho)$ is the von Neumann entropy. The  $l_{1}$ norm of coherence  is another measure which satisfies the above conditions. In this work, we mainly use the   $l_{1}$ norm of coherence  which is defined  as the sum of the absolute value of the off-diagonal elements,
\begin{equation}
    \mathcal{C}l_{1}(\rho)=\sum_{{i,j},{i\neq j}}\vert\rho_{i,j}\vert.
\end{equation}

\subsection{Time local Master equation }
The canonical form of the master equation in local time for $`d'$dimensional~\cite{hall2014canonical} is given as,

\begin{align}
 \dot{\rho}=\mathcal{L} (\rho_{t})&=\frac{-i}{h}[H(t),\rho_{t}]\nonumber \\
   &+\sum_{\mu=1}^{d^2-1}\gamma_{\mu}(t)\bigg[ \mathcal{A}_{\mu}(t)\rho_{t} \mathcal{A}_{\mu}^{\dag}(t) -\frac{1}{2}\{\mathcal{A}_{\mu}^{\dag}(t)\mathcal{A}_{\mu}(t),\rho_{t}\}\bigg]\,,
   \label{cmeq}
   \end{align}
where $\{\mathcal{A}_{\mu}(t)\}$ forms an orthonormal basis set of traceless operators, i.e; $Tr[\mathcal{A}_{\mu}(t)]=0$, $H(t)$  $Tr[\mathcal{A}_{m}^{\dag}(t)L_{n}(t)]=\delta_{mn}$, and $H(t)$ is Hermitian operator. $\gamma_{\mu}(t)$ and $L_{\mu}(t)$ are the time-dependent decoherence rates and decoherence operators, respectively. The decoherence rate $\gamma_{mu}$ is uniquely defined and invariant under the unitary transformation. The value  of $\gamma_{\mu}(t)$ determines the nature of the interaction of systems with the environment. If the decoherence rate is positive, we say that the evolution of the system under noisy interaction is time-dependent Markovian, and therefore, the quantum channel is divisible. On the other hand if $\gamma_{\mu}(t)$ is negative then the evolution is non-Markovian.\newline
The memoryless master equation of Lindblad form under Born-Markov and rotation wave approximations is, $\dot{\rho}=\frac{-i}{h}[H,\rho]
   +\sum_{\mu=1}^{d^2-1}\gamma_{\mu}\bigg[ L_{\mu}\rho L_{\mu}^{\dag} -\frac{1}{2}\{L_{\mu}^{\dag}L_{\mu},\rho\}\bigg].
   $\newline
%\subsection{Measure of non-Markovianity}
% \par~~~~~~Measure of weak non-Markovianity as a deviation from temporal self-similarity is defined as~\cite{shrikant2020},
% \begin{equation}
%   \mathcal{N}_{\mathcal{L}}=min_{\mathcal{L}^{*}}\frac{1}{T} \int_{0}^{T}\vert\vert \mathcal{L}(t)-\mathcal{L}^{*}\vert\vert dt,
%     \label{NMMSR}
%     \end{equation}
% where, $\vert\vert B\vert\vert=Tr\sqrt{B B^\dag}$ is the trace norm of the operator,  $\mathcal{L}(t)$ and $\mathcal{L}^{*}$ are the  generators at non-Markovian and Markovian regime respectively. $\mathcal{N}_{\mathcal{L}}=0$ iff the channel is quantum dynamical semigroup (QDS) and is greater than zero for a deviation from QDS. \newline  

\section{Measures of speed limit time}
Here, we discuss different measures of quantum speed limit time based on relative purity and Bures angle for open quantum systems.
\subsection{Relative purity}
A bound analogous to the MT bound based on the relative purity~\cite {del2013} for the open quantum system in which reference is made to the initial state and the dynamical map is,

\begin{equation}
    \tau_{QSL}\geq\frac{\vert \cos\theta-1\vert tr\rho_{0}^{2}}{tr[(\mathcal{L}^{\dag}\rho_{0})^2]}\geq\frac{4\theta^2tr\rho_{0}^2}{\pi^2\sqrt{tr[(\mathcal{L}^{\dag}\rho_{0})^2]}},
        \label{spdlmt}
    \end{equation}
where $\theta=\cos^{-1}[\mathcal{P}(t)]$, $\mathcal{P}(t)=tr(\rho_{t}\rho_{0})/tr(\rho_{0}^2)$ is the relative purity of initial and final states.
The generalization of time-dependent $\mathcal{L}(t)$ is 
\begin{equation}
    \tau_{QSL}\geq\frac{4\theta^2tr\rho_{0}^2}{\pi^2\overline{\sqrt{tr[(\mathcal{L}^{\dag}\rho_{0})^2]}}},
    \label{MT_RP}
\end{equation}
where $\overline{X}=\tau_{QSL}^{-1}\int_{0}^{\tau_{QSL}}X dt$.
\subsection{Bures angle}
Mandelstam-Tamm and Margolus-Levitin-types bound on speed limit time based on the geometrical distance between the initial pure state $\rho_{0}=\vert\psi_{0}\rangle\langle\psi_{0}\vert$ and final state $\rho_{\tau}$, which is a tighter bound is given as~\cite{deffner2013},
\begin{equation}
    \tau_{QSL}=\max\Bigg\{\frac{1}{\Lambda^{\textrm{op}}_{\tau}},\frac{1}{\Lambda^{\textrm{tr}}_{\tau}},\frac{1}{\Lambda^{\textrm{hs}}_{\tau}}\Bigg\} \sin^2[\mathcal{B}]
    \label{spdlmt}
\end{equation}

where $\frac{1}{\Lambda^{\textrm{op}}_{\tau}}$,$\frac{1}{\Lambda^{\textrm{tr}}_{\tau}}$, and $\frac{1}{\Lambda^{\textrm{hs}}_{\tau}}$ are operator, Hilbert-Schmidt and trace norms, respectively, Bures angle $\mathcal{B}(\rho_{0},\rho_{\tau})=\arccos\sqrt{\mathcal{F}(\rho_{0},\rho_{\tau})}$, where the Bures fidelity $\mathcal{F}(\rho_{0},\rho_{\tau})$ is $\bigg[\textrm{tr}[\sqrt{\sqrt{\rho_{0}}\rho_{\tau}\sqrt{\rho_{0}}}]\bigg]^{2}$ and,
\begin{equation}
    \Lambda^\textrm{op,tr,hs}_{\tau}=\frac{1}{\tau}\int^{\tau}_{0}dt\vert\vert \mathcal{L}(\rho_t)\vert\vert_\textrm{op,tr,hs}.
    \label{B_speed_limit}
\end{equation}
As it is known, the operators hold the following inequality 
$\vert\vert B\vert\vert_{\textrm{op}}\leq\vert\vert B\vert\vert_{\textrm{hs}}\leq\vert\vert B\vert\vert_{\textrm{tr}}$, as a result, we have, $1/\Lambda^{\textrm{op}}_{\tau}\geq1/\Lambda^{\textrm{hs}}_{\tau}\geq1/\Lambda^{\textrm{tr}}_{\tau}$, which shows that quantum speed limit time based on operator norm of the nonunitary generator provides the tighter bound on $\tau_{QSL}$ for an actual driving time $\tau$.  For analytical calculations, we use a simpler expression for an upper bound on fidelity ~\cite{miszczak2008sub}, where for density matrices $\rho_0$ and $\rho_\tau$, $\mathcal{F}(\rho_0,\rho_\tau)\leq \textrm{tr}\rho_0\rho_\tau+\sqrt{(1-\textrm{tr}\rho_0^2)(1-\textrm{tr}\rho_\tau^2)}$, with equality holding for single qubit state. In~\cite{wu2020quantum}, using this expression of upper bound on fidelity,  a modified expression of $\tau_{QSL}$ for both pure and mixed is derived and is calculated by  modifying the denominator in Eq.~\ref{spdlmt} as 
$\Lambda^{\textrm{op,tr,hs}}_{\tau}=\frac{1}{\tau}\int^{\tau}_{0}dt\vert\vert \mathcal{L}(\rho_t)\vert\vert_{\textrm{op,tr,hs}}\Bigg(1+\sqrt{\frac{1-\textrm{tr}\rho_0^2}{1-\textrm{tr}\rho_t^2}}\Bigg).
$
% \subsection{Quantum speed limit time based on Bures angle}
% \begin{equation}
%     \tau_{QSL}=\max\Bigg\{\frac{1}{\Lambda^{op}_{\tau}},\frac{1}{\Lambda^{tr}_{\tau}},\frac{1}{\Lambda^{hs}_{\tau}}\Bigg\} sin^2[\mathcal{B}]
% \end{equation}
% where $\frac{1}{\Lambda^{op}_{\tau}}$,$\frac{1}{\Lambda^{tr}_{\tau}}$, and $\frac{1}{\Lambda^{hs}_{\tau}}$ are operator, Hilbert-Schmidt and trace norms, respectively, and
% \begin{equation}
%     \Lambda^{op,tr,hs}_{\tau}=\frac{1}{\tau}\int^{\tau}_{0}\vert\vert L(\rho_t)\vert\vert_{op,tr,hs}.
% \end{equation}
% As it is known, the operators hold the following inequality 
% $\vert\vert A\vert\vert_{op}\leq\vert\vert A\vert\vert_{hs}\leq\vert\vert A\vert\vert_{tr}$, as a result, $1/\Lambda^{op}_{\tau}\geq1/\Lambda^{hs}_{\tau}\geq1/\Lambda^{tr}_{\tau}$, which shows that quantum speed limit time based on operator norm of the nonunitary generator provides the sharpest bound on $\tau_{QSL}$.
% The Bures angle $\mathcal{B}=\arccos\sqrt{\mathcal{F}(\rho_{0},\rho_{\tau})}$, where $\mathcal{F}(\rho_{0},\rho_{\tau})$ is defined as
% \begin{equation}
%     \mathcal{F}(\rho_{0},\rho_{\tau})=tr[\rho_{0}\rho_{\tau}]+\sqrt{1-tr[\rho_{0}^{2}]}\sqrt{1-tr[\rho_{\tau}^2]}.
% \end{equation}
% Equation (\ref{spdlmt}) provides sharper bounds on  quantum speed limit time for an open quantum system.
%  In this work, we investigate the dynamics of $\tau_{QSL}$ for various noisy quantum channels, and the relationship between quantum correlations and speed limit time is derived under the same conditions.
\section{The role of quantum coherence and mixedness on quantum speed limit time}
This section examines the role of quantum coherence and mixedness  on the speed of quantum evolution in various quantum processes. We calculate the speed limit time in terms of the coherence of initial and final quantum states.  Dephasing and dissipative processes are taken into consideration. We consider divisible and indivisible non-Markovian quantum maps. To this end, initially, the case of  the dephasing model (unital) is considered. 
The initial state is expressed as,
\begin{equation}
\rho_{0}=\frac{1}{2}
    \begin{pmatrix}
    1+\eta_{z} & \eta_{x}-i \eta_{y}  \\
  \eta_{x}+i \eta_{y}  & 1-\eta_{z}
\end{pmatrix},
\label{intstate}
\end{equation}
$\eta=(\eta_{x},\eta_{y},\eta_{z})$, $\eta \epsilon \mathcal{R}^3 $, and $||\eta||\leq1$.

The Quantum speed limit with respect to the coherence   in terms of relative is calculated as,
% \begin{equation}
%     \tau_{QSL}=\frac{4\sqrt{2}\cos^{-1}(\mathcal{P})^2tr\rho_{0}^2}{\pi^2/\tau{\int_{0}^{\tau}\vert\dot{p_t}p_t\sqrt{\mathcal{C}l(\rho_0)r_{x}^2+r_{y}^2}\vert dt}},
%     \label{qslt_phd}
% \end{equation}
\begin{equation}
    \tau_{QSL}=\frac{4\sqrt{2}\cos^{-1}(\mathcal{P})^2tr\rho_{0}^2}{\pi^2/\tau{\int_{0}^{\tau}\vert\frac{\dot{p_t}}{p_{t}^2}\mathcal{C}l_1(\rho_t)\vert dt}}.
    \label{qslt_dp_rp}
\end{equation}
 Relative purity in terms of initial coherence is $\mathcal{P}=(1 + p_t  \mathcal{C}l_1^2(\rho_0) + \eta_z^2)/(1 + \mathcal{C}l_1^2(\rho_0) + \eta_z^2)$.\newline
 Quantum speed limit time for non-Markovian amplitude damping (non-unital) noise is calculated as,
\begin{equation}
\small
    \tau_{QSL}=\frac{4\sqrt{2}\cos^{-1}(\mathcal{P})^2tr\rho_{0}^2}{\pi^2/\tau\int_{0}^{\tau}\vert\frac{\dot{p_t}}{p_t}\sqrt{\mathcal{C}l_1^2(\rho_0)+4(1+\eta_{z}^2)}\vert dt},
    \label{qslt_amd_rp}
\end{equation}
relative purity 
    $\mathcal{P}=(1-\eta_z + p_t\mathcal{C}l_1^2(\rho_0) + p_t \eta_z (1 + \eta_z))/(1 + \mathcal{C}l_1^2(\rho_0) + \eta_z^2)$. In both cases of dephasing and dissipative processes, we calculate the  quantum coherence of final state $\rho_{t}$, $\mathcal{C}l_{1}(\rho_{t})=p_{t}\sqrt{\eta_x^2+\eta_y^2}$. The quantum coherence and purity of the initial state are  $\mathcal{C}l_1(\rho_0)=\sqrt{\eta_x^2+\eta_y^2}$. The details of  the master equation,  time-dependent density matrices and decoherence function $p_{t}$ for  dephasing and dissipative processes are given in the Appendix.
    \newline
As mentioned, we also calculate the quantum speed limit time based on the Bures angle for both dephasing and dissipative quantum channels. The speed limit time for dephasing channel is,
% \begin{equation}
%     \tau_{QSL}=\frac{1-p_t(\mathcal{C}l_1^2(\rho_0)-\eta_z^2-l_1 l_{2 t}}{\frac{1}{\tau}\int_0^\tau dt\vert \dot{p_t}\sqrt{\eta_x^2+\eta_y^2}(1+\frac{l_1}{l_{2t}})\vert},
% \end{equation}
\begin{equation}
    \tau_{QSL}=\frac{1-p_t(\mathcal{C}l_1^2(\rho_0)-\eta_z^2-l_1 l_{2 t}}{\frac{1}{\tau}\int_0^\tau dt\vert\frac{\dot{p_t}}{p_t}\mathcal{C}l_{1}(\rho_{t})(1+\frac{l_1}{l_{2t}})\vert},
    \label{qslt_dp_bures}
\end{equation}
where we write $l_1=\sqrt{1-(\mathcal{C}l_1^2(\rho_0)+\eta_z^2)}$, and $l_{2t}=\sqrt{1-(\mathcal{C}l_1^2(\rho_t)+\eta_z^2)}$.
%where we write $l_1=\sqrt{1-(\eta_x^2+\eta_y^2+\eta_z^2)}$, and $l_{2t}=\sqrt{1-p_t^2(\eta_x^2+\eta_y^2)-\eta_z^2)}$. 
Similarly, for the quantum  dissipative process, we have,
\begin{equation}
    \tau_{QSL}=\frac{1+\eta_{z}-p_{t}(\mathcal{C}l_1^2(\rho_0)+p_{t}\eta_{z}(1+\eta_{z}))-h_{1}h_{2t}}{\frac{1}{\tau}\int_{0}^{\tau} dt \vert\dot{p_{t}} \sqrt{\mathcal{C}l_1^2(\rho_0) + 4 p_{t}^2 (1 + \eta_{z})^2}(1+\frac{h_1}{h_{2t}})\vert },
    \label{qlst_amd_bures}
\end{equation}
 we have $h_{1}=\sqrt{1-(\mathcal{C}l_1^2(\rho_0)+\eta_{z}^2)}$, and
$h_{2t}=\sqrt{p_{t}^2(2+2 \eta_{z}^2-p_{t}^2 (1+\eta_z)^2)-\mathcal{C}l_1^2(\rho(t))}$.\\
 %and $h_{1}=\sqrt{1-(\eta_{x}^2+\eta_{y}^2+\eta_{z}^2)}$, and
%$h_{2t}=\sqrt{p_{t}^2(2-\eta_{x}^2-\eta_{y}^2+2 \eta_{z}^2-p_{t}^2 (1+\eta_z)^2)}$.
Next, we show how the coherence of initial and final states influences the speed limit time of evolution under various quantum noises. We consider the initial state $\rho=\frac{1-q}{2}\textrm{I}_{2}+q \vert\chi_{\pm}\rangle\langle\chi_{\pm}\vert$ where maximally coherent pure state, $\vert\chi_{\pm}\rangle=\frac{1}{\sqrt{2}}(\vert0\rangle\pm\vert1\rangle)$.

\begin{figure}[htb]
    \centering
    \includegraphics[width=0.49\columnwidth]{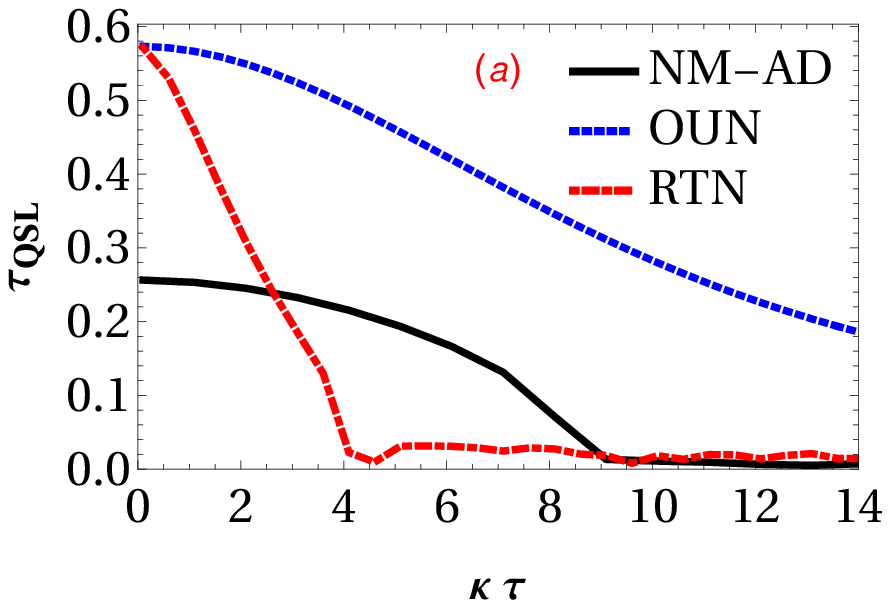}
     \includegraphics[width=0.49\columnwidth]{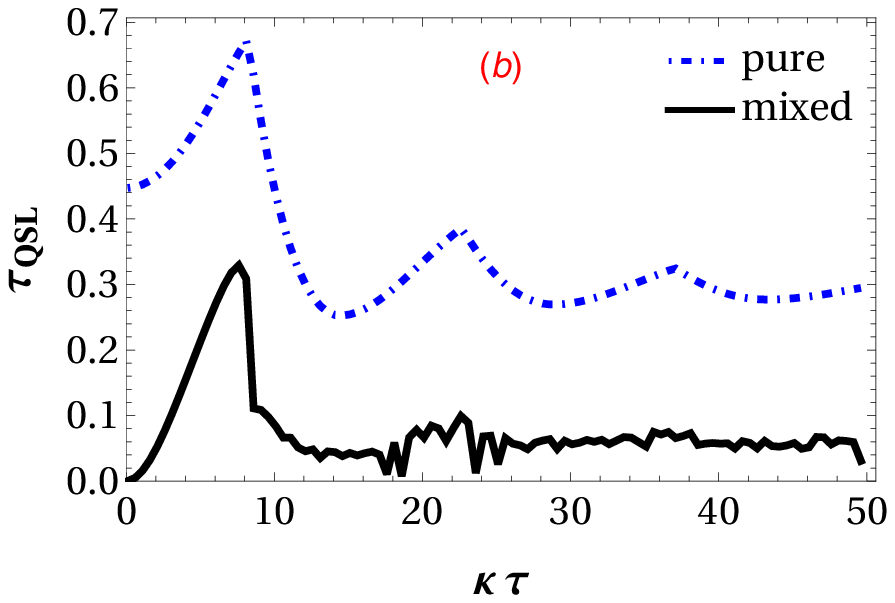}
     \includegraphics[width=0.49\columnwidth]{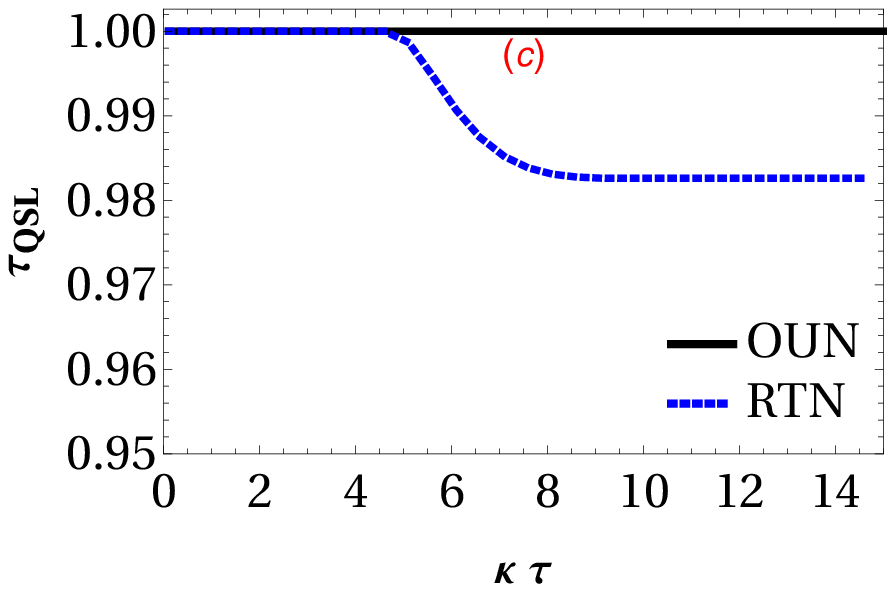}
    \includegraphics[width=0.49\columnwidth]{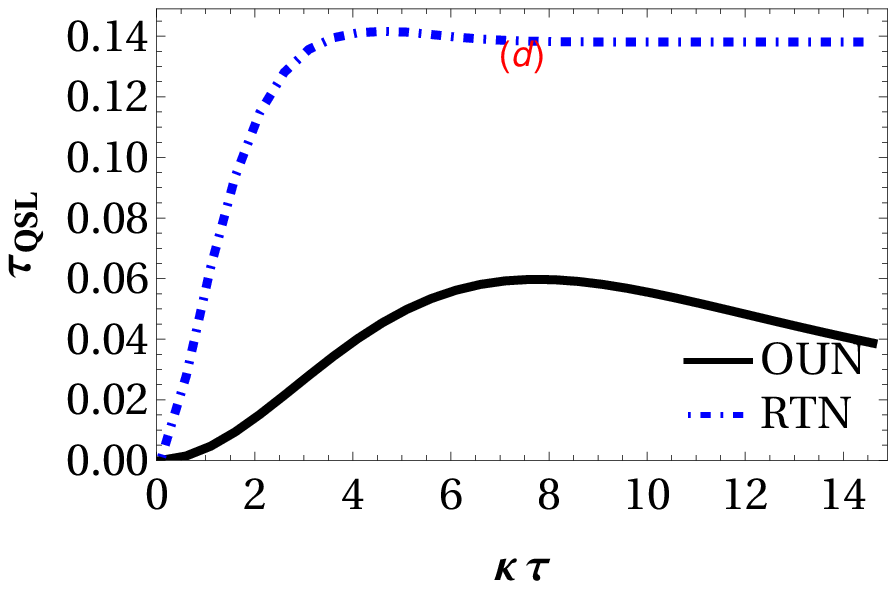}
    \caption{ Quantum speed limit time based on relative purity and fidelity is given as a function of $\kappa \tau$ for dissipative and dephasing channels for single-qubit pure $(\chi_{\pm})$ and mixed $(q=\frac{1}{2})$ coherent states. In Fig. (a), we have $\tau_{QSL}$ based on relative purity for a maximally coherent initial state. Quantum speed limit time based on Bures angle for both pure and mixed states in the case of NMAD is depicted in Fig. (b), and the cases for RTN and OUN are shown in Figs. (c) and (d), respectively. Channel parameter takes the value $\lambda=0.1\kappa$ (NMAD,OUN), $\frac{a}{\kappa}=0.6$ (RTN) and actual driving time $\tau=1$.}
    \label{one_1rp}
\end{figure}

In Eqs.~\ref{qslt_dp_rp},~\ref{qslt_amd_rp},~\ref{qslt_dp_bures} and ~\ref{qlst_amd_bures}, we have quantum speed limit time of  evolution between the states based on relative purity and Bures angle in terms of initial and final $l_1$ norm of coherence of states. In Fig.~\ref{one_1rp}(a) $\tau_{QSL}$ based on relative purity for maximally coherent pure state $\vert\chi_{\pm}\rangle$is depicted as a function of $\kappa \tau$ for all quantum noises considered in this work. Similarly, quantum speed limit time in terms of Bures angle for amplitude damping and dephasing channels for pure and mixed states $(q=\frac{1}{2})$ is shown in Fig.~\ref{one_1rp}(b) and Fig.~\ref{one_1rp}(c),~\ref{one_1rp}(d), respectively. It is clear from Eqs.~\ref{qslt_dp_rp},~\ref{qslt_amd_rp},~\ref{qslt_dp_bures},~\ref{qlst_amd_bures} and Fig.\ref{one_1rp} that the coherence and mixing of quantum states play  significant roles in determining the speed of evolution between the quantum states, especially for dissipation process. As  depicted in Fig.~\ref{one_1rp}(b), the up and down swings of speed limit time occur for quantum states with non-zero values of initial coherence. This occurs due to the trade-off between quantum coherence and mixedness under evolution.
% \begin{figure}[!htbp]
%     \centering
%     \includegraphics[height=65mm,width=1\columnwidth]{NM_VS_CL.eps}
%     \caption{Measures of coherence ($l_{1}$ norm) vs  non-Markovianity $\mathcal{N}_\mathcal{L}$ is given for NM-AD, RTN and OUN channels for maximally coherent pure  initial state ($\frac{1}{\sqrt{2}}[\vert0\rangle+\vert 1\rangle]$). In the case of NM-AD and OUN channels $\lambda=0.1\kappa$, and $\frac{c}{\kappa}=0.6$ (RTN).}
%     \label{nm_vs_cl1_nmad}
% \end{figure}

\begin{figure}
    \centering
    \includegraphics[height=65mm,width=1\columnwidth]{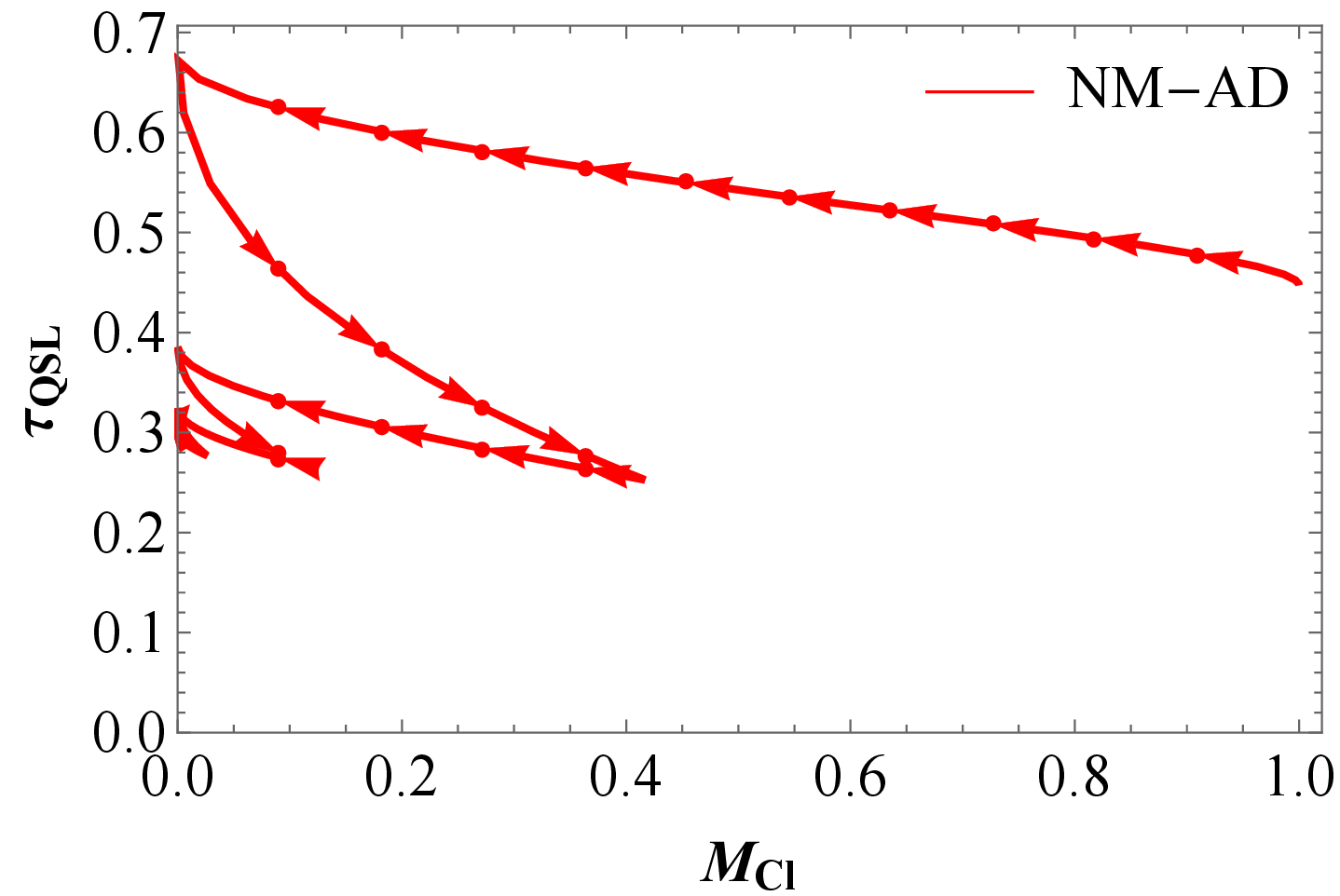}
    \caption{The parametric trajectory of quantum speed limit time as a function of $M_{\mathcal{C}l}$ is depicted for the maximally coherent pure state in the dissipative process (NMAD channel), a method based on the relative purity is availed to calculate the speed limit time. Revival of quantum coherence and speed limit time occurs due to the effect of quantum non-Markovianity. The interplay between coherence, mixedness, and speed limit time is given for  $\lambda=0.1\kappa$, and actual driving time $\tau=1$. }
    \label{path_b_mc_amd}
\end{figure}

In~\cite{singh2015maximally} connection between quantum coherence and mixedness of quantum states is established. It is shown that for any arbitrary density matrix $\rho$ of dimension $d$, the amount of coherence $\mathcal{C}l_1$ is restricted by the quantity of mixedness $S_l$ by the inequality\begin{equation}
    M_{\mathcal{C}l}=\frac{\mathcal{C}{l}_1^2(\rho)}{(d-1)^2}+S_l(\rho)\leq1.
    \label{mcl}
\end{equation} 
% The rate of $M_{\mathcal{C}l}$ is 
% \begin{equation}
%     \frac{dM_{\mathcal{C}l}}{dt}=\frac{2\mathcal{C}l_1\dot{\mathcal{C}l_1}+d(d-1)(1-2\textrm{tr}(\rho_t\mathcal{L}\rho_t))}{(d-1)^2}.
% \end{equation}
The linear entropy $S_l(\rho)=\frac{d}{d-1}(1-\textrm{tr}\rho^2)$ gives the mixedness of the quantum state $\rho$ of dimension $d$. For a maximally coherent state to a fixed mixedness, we have $M_{\mathcal{C}l}=1$. In this work, we investigate the interplay between the coherence and mixedness of quantum states under non-unitary evolution and its impacts on quantum speed limit time for multi-qubit states.\newline
The mixedness of a general single qubit's state under the dephasing process is estimated as $1-p_{t}(\eta_{x}^2+\eta_{y}^2)-\eta_{z}^2$, and coherence as aforementioned is $\mathcal{C}{l}_1=p_{t}\sqrt{\eta_{x}^2+\eta_{y}^2}$. Plugging the expressions of coherence and mixedness in  Eq.\ref{mcl} gives $M_{\mathcal{C}l}=1-\eta_{z}^2$, which is invariant throughout the evolution of quantum states. For example, $M_{\mathcal{C}l}$ of maximally coherent $\vert\chi_{\pm}\rangle$ is equal to $1$, is invariant under dephasing process; it is not so under a dissipative process.
\newline For a dissipative process, we have  linear entropy $S_{l}(\rho_t)=p_t^2(2-(\eta_x^2-\eta_y^2+2\eta_z-p_t^2(1+\eta_z)^2))$, coherence as mentioned before, which together gives,
\begin{equation}
    M_{\mathcal{C}l}=p_{t}^2(1+\eta_{z})(2-p_{t}^2(1+\eta_{z})).
    \label{mcl_amd}
\end{equation}
From Eq.~\ref{mcl_amd}, it is evident that $M_{\mathcal{C}l}$ is not invariant under amplitude damping noise, which along with the initial coherence and mixedness, determine the nature of $\tau_{QSL}$. To analyse this in detail, we consider the case of the maximally mixed coherent initial state ($M_{\mathcal{C}l}=1$) $\rho=\frac{1-q}{2}\textrm{I}_2+q \vert\chi_{\pm}\rangle\langle\chi_{\pm}\vert$. In Fig.~\ref{path_b_mc_amd}, the parametric trajectory of quantum speed limit time is depicted as a function of $M_{\mathcal{C}l}$ for an initial maximally coherent pure state ($q=1$). The path of evolution from initial to final states characterizes the nature of the dissipative process. Multiple values of speed limit time for a fixed $M_{\mathcal{C}l}$ reveal the trade-off between coherence and mixedness and the existence of backflow of information from the reservoir to the system. For a mixed initial state, the evolution of the quantum state takes a different trajectory which sheds light on identifying $\tau_{QSL}$ as a dynamical witness to distinguish quantum states. This property of quantum speed limit time is discussed in the forthcoming section.

\section{Multi-qubit system: coherence, mixedness and distinguishabilty}
A state of an $\textrm{N}-$dimensional quantum system can be expressed in terms of generators $\hat{\Gamma}_i$ of SU(N), as
\begin{equation}
    \rho=\frac{\textrm{I}}{\textrm{N}} +1/2\sum_{i=1}^{\textrm{N}^2-1} r_i\hat{\Gamma}_i,
\end{equation}
where we have $r_i=\textrm{Tr}[\rho \hat{\Gamma_i}]$. The generators $\hat{\Gamma_{i=1...N^2-1}}$ satisfy the conditions (1) $\hat{\Gamma_i}=\hat{\Gamma}_i^{\dag}$, (2) $\textrm{Tr}(\hat{\Gamma}_i)=0$ and (3) $\textrm{Tr}(\hat{\Gamma}_i\hat{\Gamma}_j)=2\delta_{ij} $. The positivity condition of the density matrix is stated in terms of the characteristic equation for $\rho$. The state is positive semidefinite iff all the coefficients of the polynomial $\textrm{det}(\lambda I-\rho)=\sum_{i=0}^\textrm{N} (-1)^i \textrm{B}_i\lambda^{\textrm{N}-1}=0$ are such that $\textrm{B}_i\geq 1$ for $1\leq i\leq \textrm{N}$ and $\textrm{B}_0=1$.\newline
In this section, we consider different initial  states, which mainly fall in a class of  $\textrm{X}$ states. Along with that, we consider the cases of maximally coherent entangled pure and mixed states. We investigate the effect of  the trade-off between coherence and mixedness in determining the speed limit time. \newline
A general two-qubit density matrix is written as,
\begin{equation}
    \rho_{AB}=\frac{1}{4}(I_4+\mathbf{r}.\sigma\otimes I+\mathbf{s}.I\otimes\sigma+\sum_{i,j}k_{i,j}\sigma_{i}\otimes\sigma_{j}),
\end{equation}
 where $\mathbf{r}$ and $\mathbf{s}$ are vectors and $\sigma_{i}^{'s}$ are the Pauli's matrices. $\rho_{AB}$ reduces to  Bell-diagonal state, a class of X state for $\mathbf{r}=0=\mathbf{s}$ such that  $\rho_{X_{B}}=\frac{1}{4}(I_{4}+\sum_{i}k_{i}\sigma_{i}\otimes\sigma_i)$. The Bell diagonal density matrix is written as,
 \begin{equation}
     \rho_{X_{B}}=\frac{1}{4}\left(
\begin{array}{cccc}
 1+k_3 & 0 & 0 & k_1-k_2 \\
 0 & 1-k_3 & k_1+k_2 & 0 \\
 0 & k_1+k_2 & 1-k_3 & 0 \\
 k_1-k_2 & 0 & 0 & 1+k_3 \\
\end{array}
\right).
 \end{equation}
Initial coherence and purity of the state $\rho_{X_{B}}$ are $\frac{1}{2}(\vert k_1-k_2\vert+\vert k_1+k_2\vert)$ and $\frac{1}{4}(k_1^2+k_2^2+k_3^2+1)$, respectively.
The time-dependent Bell diagonal density matrix under dephasing noise is,
\begin{align*}
%\begin{equation}
\small
\rho_{X_{B}}(t)=\frac{1}{4}\left(
\scriptstyle
\begin{array}{cccc}
 1+k_3& 0 & 0 &p_t^2(k_1-k_2) \\
 0 & 1-k_3 & p_t^2 (k_1+k_2) & 0 \\
 0 & p_t^2 (k_1+k_2) & 1-k_3 & 0 \\
 p_t^2 (k_1-k_2) & 0 & 0 & 1+k_3 \\
\end{array}
\right).
% \end{equation}
 \end{align*}
We calculate coherence $ Cl_{1}(\rho_{X_{B}}(t))=\frac{1}{2}p_{t}^2(\vert  k_{1}-k_{2}\vert+\vert  k_{1}+k_{2}\vert)$ and purity
 $\textrm{tr}(\rho_{X_{B}}(t)^2)=\frac{1}{4} \left(1+p_t^4 \left(k_1^2+k_2^2\right)+k_3^2\right))$. Quantum speed limit time based on relative purity is,
 \begin{equation}   
\tau_{QSL}=\frac{4(\cos^{-1}(\mathcal{P}))^2tr\rho_{X_{B}}(0)^2}{\pi^2/\tau \int_0^{\tau}dt{ p_t\dot{p_t}\sqrt{k_{1}^2+k_{2}^2}}},
    \label{two_tqsl_rp__dp}
\end{equation}
 where relative purity $\mathcal{P}=\frac{1+k_3^2+\left(k_1^2+k_2^2\right) p_t^2}{1+k_1^2+k_2^2+k_3^2}$. The trade-off between coherence and mixedness of quantum states under the dephasing process is,
 \begin{equation}
     M_{\mathcal{C}l}=\frac{1}{18} \left(-5 \left(k_1^2+k_2^2\right) p_t^4+p_t^4\vert k_1^2-k_2^2\vert-6 k_3^2+18\right).
     \label{mcl_two_dp}
 \end{equation}
 The time-dependent density matrix under the dissipative process is written as,
 \begin{widetext}
 \small
  \begin{equation}
     \rho_{X_{B}}(t)=\left(
\begin{array}{cccc}
 1-p_t^2+\frac{1}{4} \left(1+k_3\right) p_t^4 & 0 & 0 & \frac{1}{4} \left(k_1-k_2\right) p_t^2 \\
 0 & \frac{1}{4} \left(2 p_t^2-\left(1+k_3\right) p_t^4\right) & \frac{1}{4} \left(k_1+k_2\right) p_t^2 & 0 \\
 0 & \frac{1}{4} \left(k_1+k_2\right) p_t^2 & \frac{1}{4} \left(2 p_t^2-\left(1+k_3\right) p_t^4\right) & 0 \\
 \frac{1}{4} \left(k_1-k_2\right) p_t^2 & 0 & 0 & \frac{1}{4} \left(1+k_3\right) p_t^4 \\
\end{array}
\right).
 \end{equation}
  \end{widetext}
Quantum speed limit time based on relative purity for Bell-diagonal state is estimated as,
\begin{equation}   
\tau_{QSL}=\frac{4(\cos^{-1}(\mathcal{P}))^2tr\rho_{X_{B}}(0)^2}{\pi^2/\tau \int_0^{\tau}dt{ \frac{\dot{p_t}}{p_t}\sqrt{2+k_{1}^2+k_{2}^2+4k_{3}^2}}}.
    \label{two_tqsl_rp__NMAD}
\end{equation}
We have relative purity 
\begin{equation*}
\mathcal{P}=\frac{1+k_3+p_t^2 \left(k_1^2+k_2^2+k_3 \left(-2+\left(1+k_3\right) p_t^2\right)\right)}{1+k_1^2+k_2^2+k_3^2}.
\end{equation*}
As for the single-qubit state, we also calculate the speed limit time for the two-qubit Bell diagonal state based on Bures angle. $\tau_{QSL}$ for dephasing process is,
\begin{widetext}
\begin{equation}
    \tau_{QSL}=\frac{\frac{1}{4} \left(3-k_3^2-\left(k_1^2+k_2^2\right) p_t^2-\sqrt{\left(-3+k_1^2+k_2^2+k_3^2\right)
   \left(-3+k_3^2+\left(k_1^2+k_2^2\right) p_t^4\right)}\right)}{\frac{1}{\tau}\int^{\tau}_{0}dt\max \{\frac{1}{4}\dot{p_t}(k_1-k_2) p_t^3,\frac{1}{4} \dot{p_t}(k_1+k_2) p_t^3\}(1-\sqrt{\frac{k_1^2+k_2^2+k_3^2-3}{\left(k_1^2+k_2^2\right) p_t^4+k_3^2-3}})}.
\end{equation}
\end{widetext}
Similarly, analytical expression of  $\tau_{QSL}$  for dissipative process can be shown to be,
\begin{widetext}
\small
\begin{equation}
    \tau_{QSL}=\frac{\frac{1}{4} \left(2 k_3-k_1^2-k_2^2\right) p_t^2-\frac{1}{4} \left(k_3+k_3^2\right) p_t^4+\frac{1}{4}
   \left(3-k_3-q \sqrt{\left(\sum _i k_i^2-3\right) p_t^2 \left(-8+\left(8+k_1^2+k_2^2+2 k_3\right) p_t^2-4 \left(1+k_3\right)
   p_t^4+\left(1+k_3\right){}^2 p_t^6\right)}\right)}{\frac{1}{\tau}\int^{\tau}_{0}dt \max\{\sqrt{\zeta_1},\sqrt{\zeta_2},\sqrt{\zeta_3},\sqrt{\zeta_4}\} (1-\sqrt{\frac{k_1^2+k_2^2+k_3^2-3}{p_t^2 \left(p_t^2 \left(\left(k_3+1\right) p_t^2 \left(\left(k_3+1\right)
   p_t^2-4\right)+k_1^2+k_2^2+2 k_3+8\right)-8\right)}})}.
   \label{speed_two_B_NMAD}
\end{equation}
To investigate the interplay between coherence and mixedness, we have,
\begin{equation}
    M_{\mathcal{C}l}=\frac{1}{18} \left(-6 \left(k_3+1\right){}^2 p_t^8+24 \left(k_3+1\right) p_t^6+\left(-5 k_1^2-5 k_2^2-12
   \left(k_3+4\right)+\vert k_1^2-k_2^2\vert\right) p_t^4+48 p_t^2\right).
   \label{mcl_two_nmad}
\end{equation}
\end{widetext}
 In Eq.~\ref{speed_two_B_NMAD} we have,
\begin{align*}
\zeta_1&=\frac{1}{4} \dot{p_t}{}^2 p_t^2 (-2 (k_3+1) p_t^2+k_1+k_2+2){}^2\\
\zeta_2&=\frac{1}{4} \dot{p_t}{}^2 p_t^2 (2(k_3+1) p_t^2+k_1+k_2-2){}^2\\
\zeta_3&=\frac{1}{4} \dot{p_t}{}^2 p_t^2(4 (k_3+1) p_t^2 ((k_3+1) p_t^2-2)\\
&-4 \sqrt{k_1-k_2){}^2+4}
   ((k_3+1) p_t^2-1)+(k_1-k_2){}^2+8)\\
\zeta_4&=\frac{1}{4} \dot{p_t}{}^2 p_t^2(4 (k_3+1) p_t^2 ((k_3+1) p_t^2-2)\\
&+4 \sqrt{k_1-k_2){}^2+4}
   ((k_3+1) p_t^2-1)+(k_1-k_2){}^2+8).
\end{align*}

The role of coherence and mixedness on quantum speed limit time is now investigated. We mainly consider the cases of maximally entangled Bell states, Werner states, and maximally coherent entangled pure and mixed states. Maximally entangled Bell states are,
\begin{eqnarray}
\nonumber
   \vert \phi_{\pm}\rangle=\frac{1}{\sqrt{2}}(\vert00\rangle\pm\vert11\rangle)\\
   \vert \psi_{\pm}\rangle=\frac{1}{\sqrt{2}}(\vert01\rangle\pm\vert10\rangle).
   \label{two_ghz_basis}
\end{eqnarray}
\begin{figure}[!htb]
    \centering
    \includegraphics[width=0.49\columnwidth]{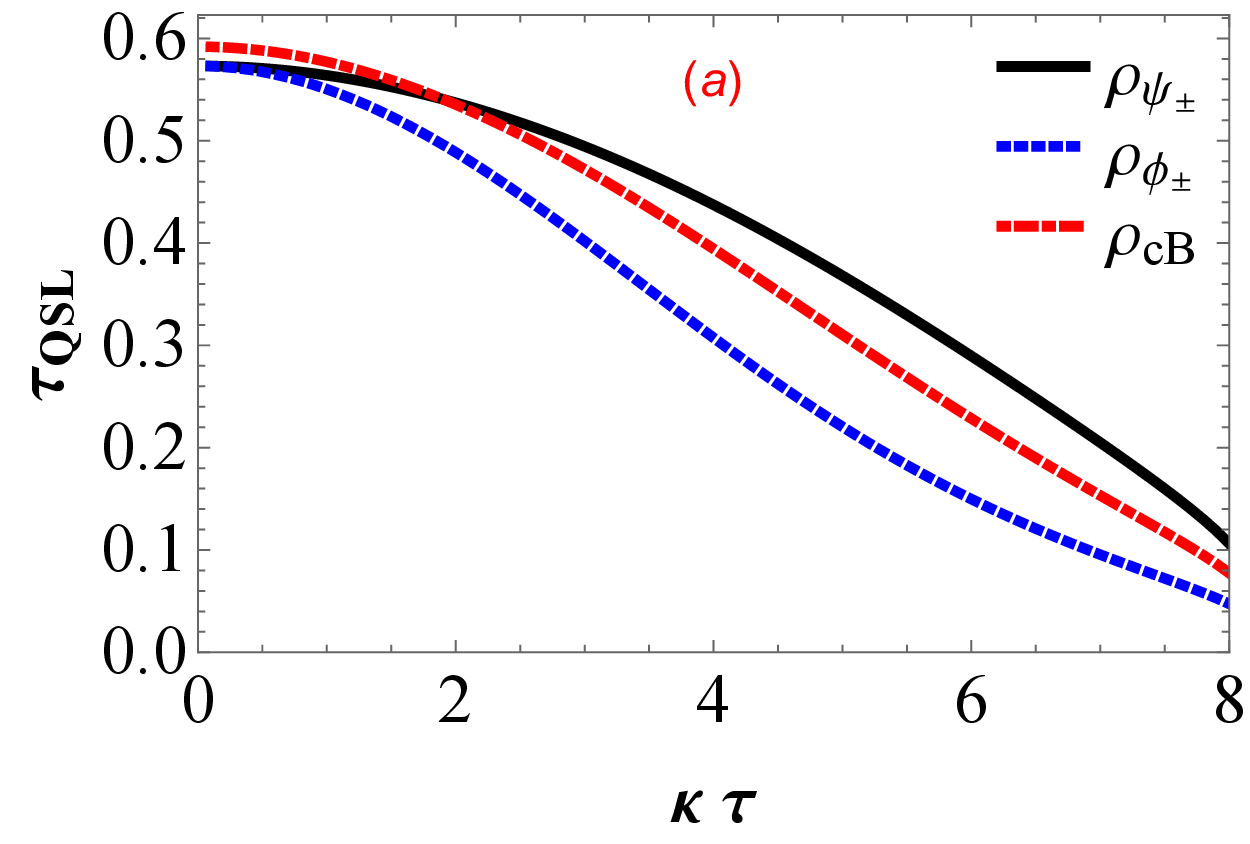}
     \includegraphics[width=0.49\columnwidth]{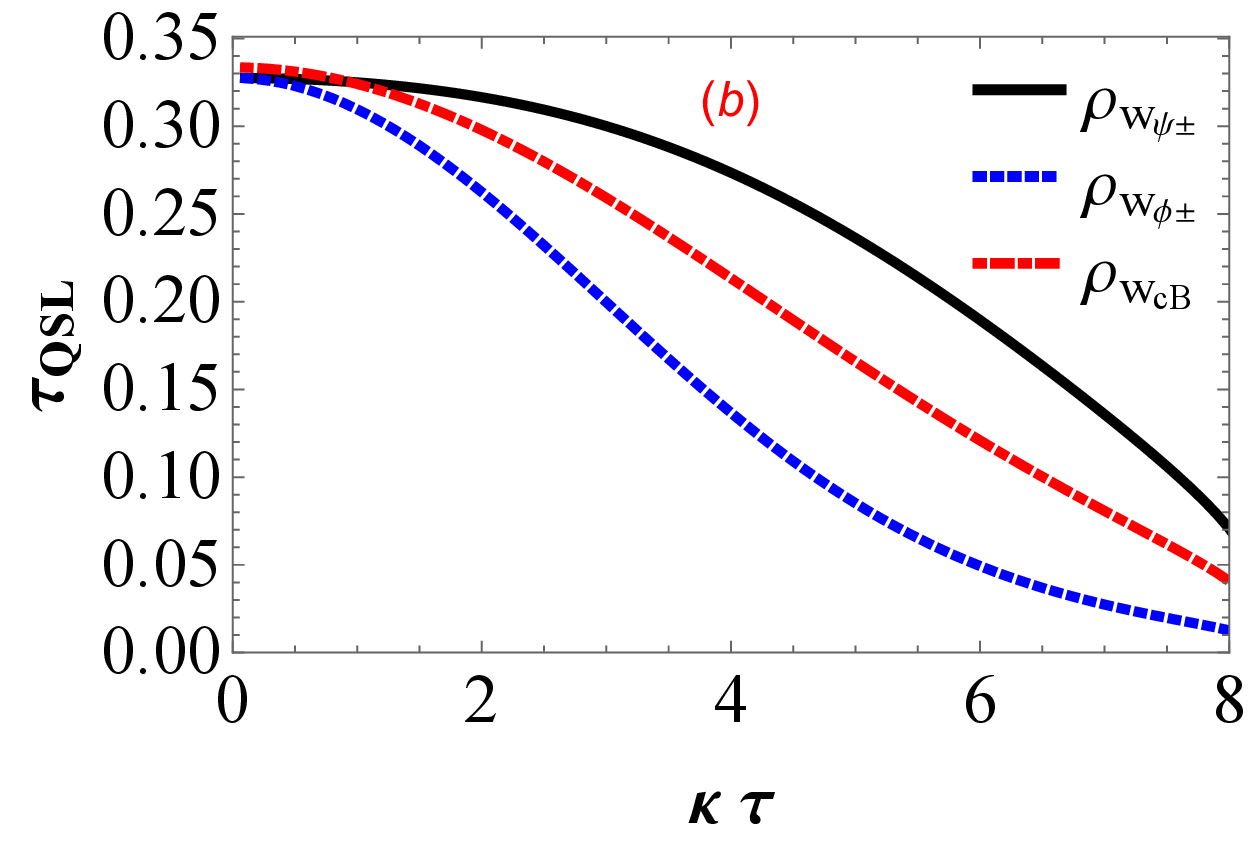}
     \includegraphics[width=0.49\columnwidth]{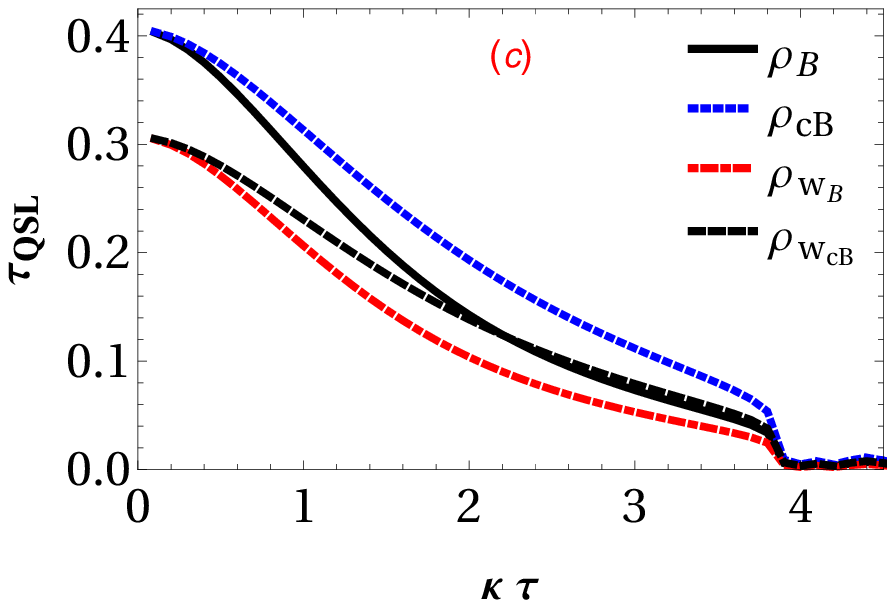}
    \includegraphics[width=0.49\columnwidth]{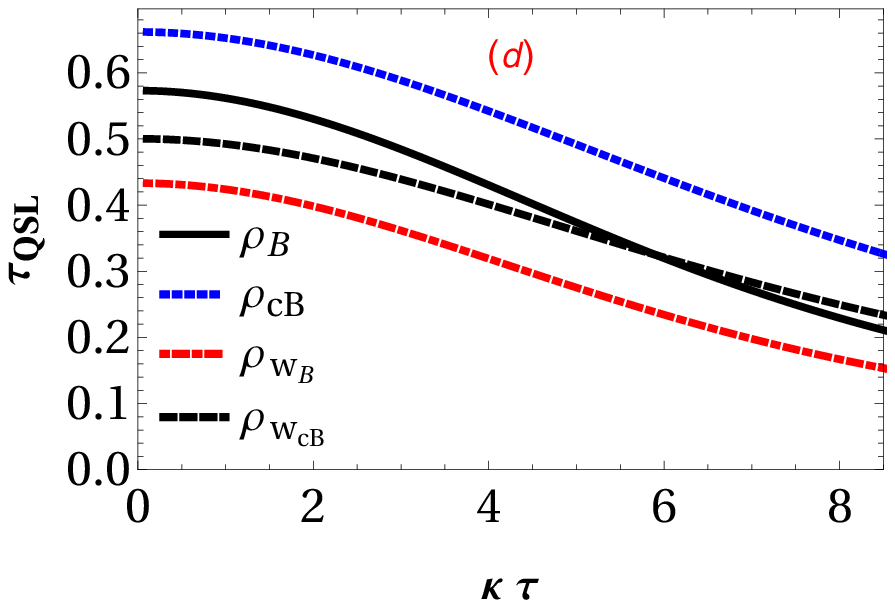}
    \caption{ Quantum speed limit time based on relative purity is depicted as a function of $\kappa \tau$ for dissipative and dephasing channels for two-qubit entangled Bell, Werner,  maximally coherent entangled, maximally coherent mixed entangled initial states. Figs (a)-(b), we have the speed limit time for pure and mixed $(q=\frac{1}{2})$ entangled states for the NMAD channel, and the cases of RTN and OUN channels are shown in Figs. (c) and (d), respectively. Channel parameter takes the value $\lambda=0.1\kappa$ (NMAD, OUN),$\frac{a}{\kappa}=0.6$ (RTN), and actual driving time $\tau=1$.}
    \label{two_1rp}
\end{figure}

Maximally coherent entangled states $\rho_{cB}$ can be obtained from maximally entangled Bell states by the action of unitary operator $U$,
\begin{equation}
U=\frac{1}{\sqrt{2}}
    \begin{pmatrix}
    1&0&1&0\\
    0&1&0&1\\
    1&0&-1&0\\
    0&1&0&-1
    \end{pmatrix}.
\end{equation}
Similar  unitary matrices for multi-qubit states can also be identified. Using Eq.~\ref{two_tqsl_rp__dp} and \ref{two_tqsl_rp__NMAD}, in Fig.~\ref{two_1rp}, we have quantum speed limit time based on relative purity  as a function of $\kappa\tau$ for entangled coherent pure and mixed states. The cases are checked out for both CP-(in)divisible channels. We have mixed entangled state $\rho_{W_{B}}=\frac{1-q}{4}\textrm{I}_4+q\vert\mathcal{B}\rangle\langle\mathcal{B}\vert$, where $\mathcal{B}$ is any of the maximally entangled Bell states. Similarly, we construct maximally  coherent entangled mixed states $\rho_{W_{cB}}$ by replacing the Bell states with maximally coherent entangled  states of the Werner state. In Fig.~\ref{two_1rp} (a)-(b), we have non-Markovian amplitude damping channel, while Fig.~\ref{two_1rp}(c) and Fig.~\ref{two_1rp}(d) depict the  RTN and OUN dephasing channels, respectively. As it's seen from Figs.~\ref{two_1rp} (a) and (b), for the dissipative process, the behaviour of $\tau_{QSL}$ is different for Bell states $(\phi_{\pm}$ and $\psi_{\pm})$, and maximally coherent Bell states $\rho_{cB}$. An increase of coherence does not increase, on average,  the minimum time of evolution between initial and fixed final states as one expects for the dissipation process, and the nature of $\tau_{QSL}$ for all maximally coherent entangled states is identical. An increase in coherence increases speed limit time between quantum states under dephasing (Figs.~\ref{two_1rp} (c) and (d)).  From Figs.~\ref{two_1rp} it's clear that the behaviour of speed limit time is distinct for different Bell states, which brings out the  possibility of using $\tau_{QSL}$ to distinguish quantum states. Further discussion on this is made in the next section.\newline
 
\begin{figure}
    \centering
    \includegraphics[height=65mm,width=1\columnwidth]{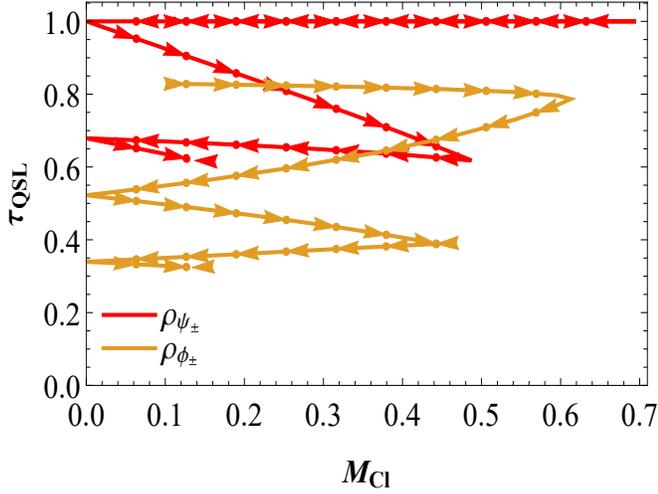}
    \caption{The parametric trajectory of quantum speed limit time as a function of $M_{\mathcal{C}l}$ is depicted for maximally entangled Bell states for the dissipative process (NMAD). The speed limit time is estimated in terms  of Bures angle. Revival of quantum coherence and speed limit time occurs due to quantum non-Markovianity of quantum noise. The interplay between coherence, mixedness, and speed limit time is given for  $\lambda=0.1\kappa$, and actual driving time $\tau=1$. }
    \label{path_2B_bell_amd}
\end{figure}
\begin{figure}
    \centering
    \includegraphics[height=65mm,width=1\columnwidth]{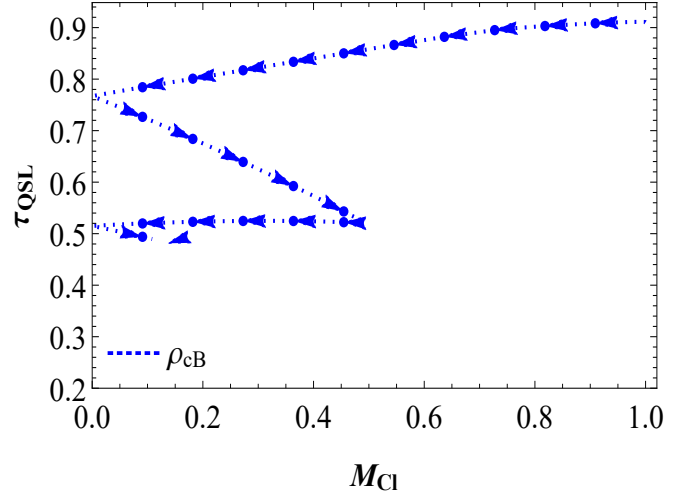}
    \caption{The parametric trajectory of quantum speed limit time as a function of $M_{\mathcal{C}l}$ is depicted for a maximally coherent entangled two-qubit state for the dissipative process (NMAD). Bures angle measure is availed to calculate the speed limit time. Left and right swings of quantum coherence and speed limit time occur due to quantum non-Markovianity. The interplay between coherence, mixedness, and speed limit time is given for  $\lambda=0.1\kappa$, and actual driving time $\tau=1$.}
    \label{path_2B_mcb_amd}
\end{figure}
\begin{figure}
    \centering
    \includegraphics[height=65mm,width=1\columnwidth]{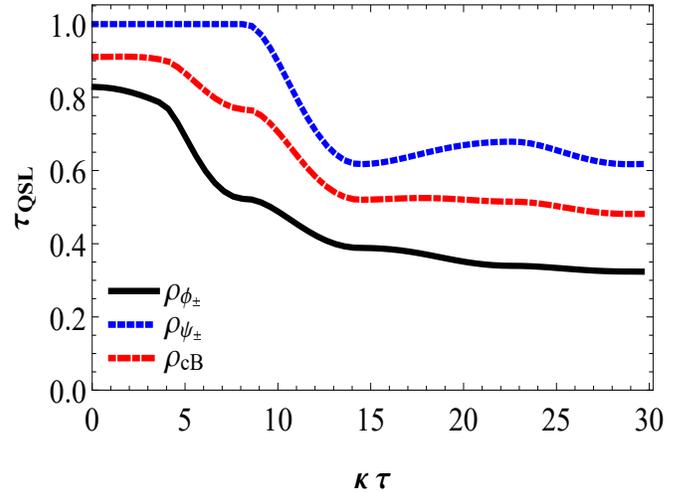}
    \caption{Quantum speed limit time in terms of Bures angle for maximally entangled states and maximally coherent entangled state is plotted as a function of $\kappa\tau$ for NMAD channel. Channel parameter takes the value $\lambda=0.1\kappa$, and actual driving time $\tau=1$. }
    \label{two_B_CB_NMAD}
\end{figure}
\begin{figure}[!htb]
    \centering
    \includegraphics[height=65mm,width=1\columnwidth]{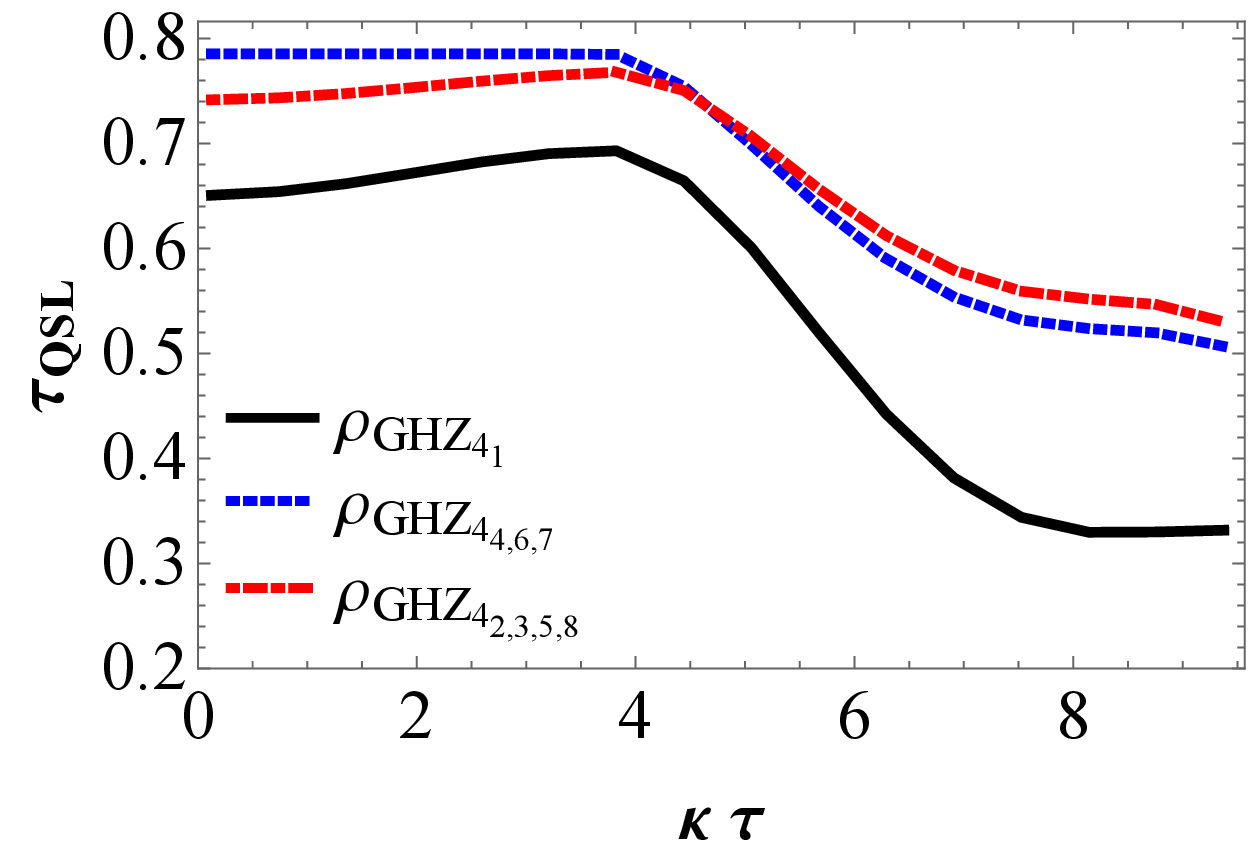}
    \caption{Quantum speed limit time in terms of Bures angle for four qubits maximally entangled states as a function of $\kappa\tau$ for NMAD channel. Channel parameter takes the value $\lambda=0.1\kappa$, and actual driving time $\tau=\frac{\pi}{4}$. }
    \label{four_CB}
\end{figure}

The trade-off between the  mixedness and coherence for two-qubit states under pure dephasing and dissipative processes is analyzed using Eq.~\ref{mcl_two_dp} and \ref{mcl_two_nmad}. The behaviour of $\tau_{QSL}$ on the  complementarity between mixedness and coherence  is significant  under dissipative conditions. In Figs.~\ref{path_2B_bell_amd} and \ref{path_2B_mcb_amd}, $\tau_{QSL}$ in terms of Bures measure is depicted as a function of $M_{\mathcal{C}l}$ for maximally entangled Bell states and maximally coherent entangled state $\rho_{cB}$, respectively. From Fig.~\ref{path_2B_bell_amd}, it is clear that two Bell states $\psi_{\pm},\phi_{\pm}$ take two different parametric trajectories of quantum speed time, which helps to distinguish them. The left and right swings of the parametric trajectory occurs due to the trade-off between mixedness and coherence and is a signature of information backflow in the non-Markovian process. Different from Bell states, for maximally coherent entangled states, we have initial $M_{\mathcal{C}l}=1$ (Fig.~\ref{path_2B_mcb_amd}). In this case (Fig.~\ref{path_2B_mcb_amd}), all maximally coherent entangled states take the same parametric trajectory of speed limit time. Fig.~\ref{two_B_CB_NMAD} complements Fig.~\ref{two_1rp}(a) for $\tau_{QSL}$ computed in terms on Bures angle.\newline
 Proceeding in this manner, we now consider three-qubit GHZ entangled states whose basis is given as,
\begin{eqnarray}
\nonumber
\vert GHZ_{3,_1}^{\pm}\rangle=\frac{1}{\sqrt{2}}[\vert 000\rangle\pm\vert 111\rangle]\\ \nonumber
\vert GHZ_{3,_2}^{\pm}\rangle=\frac{1}{\sqrt{2}}[\vert 001\rangle\pm\vert 110\rangle]\\ \nonumber
\vert GHZ_{3,_3}^{\pm}\rangle=\frac{1}{\sqrt{2}}[\vert 010\rangle\pm\vert 101\rangle]\\
\vert GHZ_{3,_4}^{\pm}\rangle=\frac{1}{\sqrt{2}}[\vert 011\rangle\pm\vert 100\rangle].
\label{three_ghz_basis}
\end{eqnarray}
% Three-qubit mixed entangled state is written as,
% \begin{equation}
%     \rho_{WGHZ}=\frac{(1-p)}{d}I_d+p\vert GHZ\rangle\langle GHZ\vert
%\end{equation}
% \begin{figure}[htb]
%     \centering
%     \includegraphics[width=0.49\columnwidth]{three_1rp_CB_NMAD.eps}
%      \includegraphics[width=0.49\columnwidth]{three_1rp_WCB_NMAD.eps}
%      \includegraphics[width=0.49\columnwidth]{three_1rp_CB_RTN.eps}
%     \includegraphics[width=0.49\columnwidth]{three_1rp_CB_OUN.eps}
%     \caption{ Quantum speed limit time based on relative purity is plotted as a function of $\kappa \tau$ for dissipative and dephasing channels for three-qubit entangled GHZ, Werner,  maximally coherent entangled, maximally coherent mixed entangled initial states. Figs (a)-(b), we have a speed limit time for pure and mixed entangled states for the NMAD channel, and the cases of RTN and OUN channels are depicted in Figs. (c) and (d), respectively. Channel parameter takes the value $\lambda=0.1\kappa$(NMAD and OUN),$\frac{a}{\kappa}=0.6$ (RTN), and actual driving time $\tau=\pi/4$.}
%     \label{three_1rp}
% \end{figure}
\begin{figure}
    \centering
    \includegraphics[height=65mm,width=1\columnwidth]{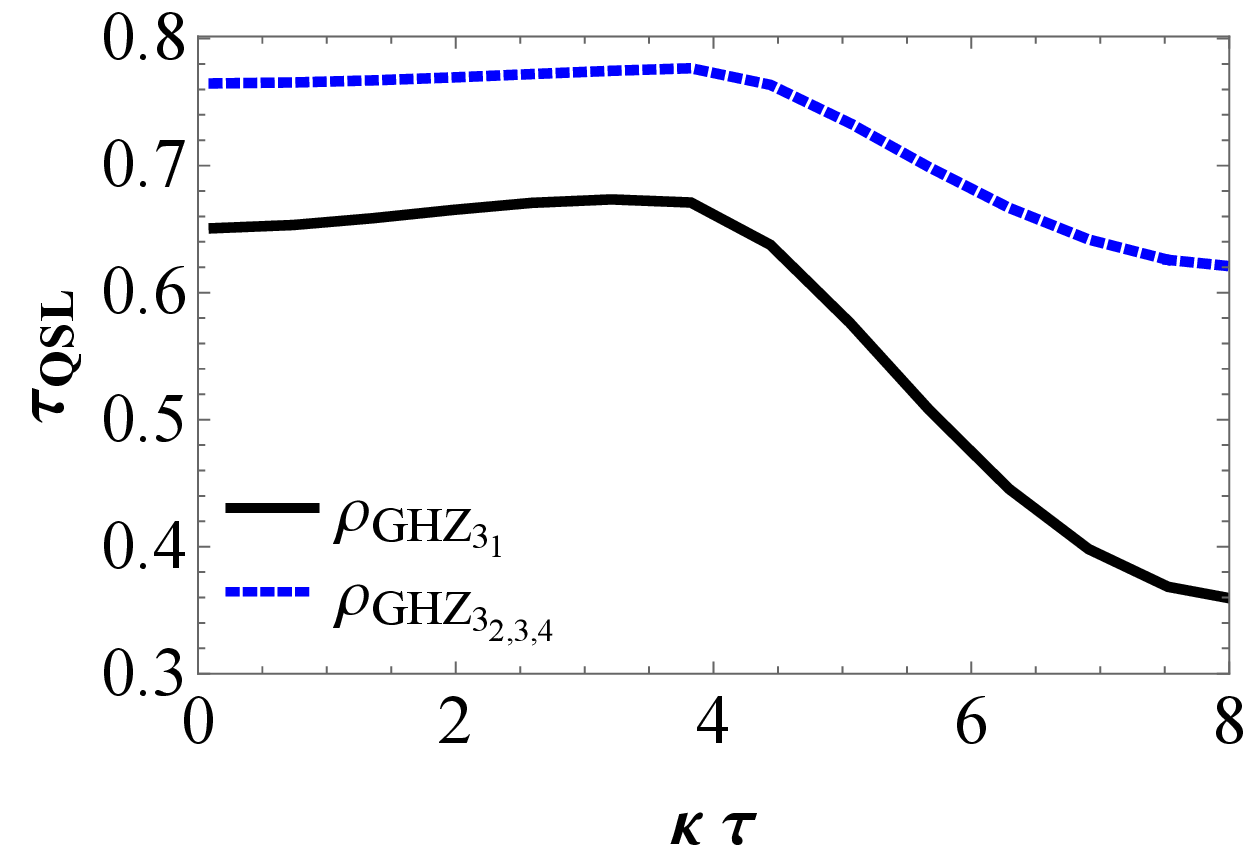}
    \caption{Quantum speed limit time in terms of Bures angle for three-qubit maximally entangled states for NMAD channel. Channel parameter takes the value $\lambda=0.1\kappa$, and actual driving time $\tau=\frac{\pi}{4}$. }
    \label{three_CB}
\end{figure} 
Similarly, the four-qubit maximally entangled GHZ basis is written as,
\begin{eqnarray}
\nonumber
\vert GHZ_{4,_1}^{\pm}\rangle=\frac{1}{\sqrt{2}}[\vert 0000\rangle\pm\vert 1111\rangle]\\ \nonumber
\vert GHZ_{4,_2}^{\pm}\rangle=\frac{1}{\sqrt{2}}[\vert 0001\rangle\pm\vert 1110\rangle]\\ \nonumber
\vert GHZ_{4,_3}^{\pm}\rangle=\frac{1}{\sqrt{2}}[\vert 0010\rangle\pm\vert 1101\rangle]\\ \nonumber
\vert GHZ_{4,_4}^{\pm}\rangle=\frac{1}{\sqrt{2}}[\vert 0011\rangle\pm\vert 1100\rangle]\\ \nonumber
\vert GHZ_{4,_5}^{\pm}\rangle=\frac{1}{\sqrt{2}}[\vert 0100\rangle\pm\vert 1011\rangle]\\ \nonumber
\vert GHZ_{4,_6}^{\pm}\rangle=\frac{1}{\sqrt{2}}[\vert 0101\rangle\pm\vert 1010\rangle]\\ \nonumber
\vert GHZ_{4,_7}^{\pm}\rangle=\frac{1}{\sqrt{2}}[\vert 0110\rangle\pm\vert 1001\rangle]\\
\vert GHZ_{4,_8}^{\pm}\rangle=\frac{1}{\sqrt{2}}[\vert 0111\rangle\pm\vert 1000\rangle].
\label{four_ghz_basis}
\end{eqnarray}
 Figs.~\ref{three_CB} and~\ref{four_CB} depict the speed limit time as a function of $\kappa \tau$ for all entangled three and four-qubit GHZ states. The usefulness of $\tau_{QSL}$ as a witness to distinguish quantum states are discussed in detail below.
% \begin{figure}[htb]
%     \centering
%     \includegraphics[width=0.49\columnwidth]{four_1rp_CB_NMAD.eps}
%      \includegraphics[width=0.49\columnwidth]{four_1rp_WCB_NMAD.eps}
%      \includegraphics[width=0.49\columnwidth]{four_1rp_CB_RTN.eps}
%     \includegraphics[width=0.49\columnwidth]{four_1rp_CB_OUN.eps}
%     \caption{ Quantum speed limit time based on relative purity is plotted as a function of $\kappa \tau$ for dissipative and dephasing channels for four- qubit entangled GHZ, Werner,  maximally coherent entangled, maximally coherent mixed entangled initial states. Figs (a)-(b), we have a speed limit time for pure and mixed entangled states for the NMAD channel, and the cases of RTN and OUN channels are depicted in Figs. (c) and (d), respectively. Channel parameter takes the value $\lambda=0.1\kappa$(NMAD and OUN),$\frac{a}{\kappa}=0.6$ (RTN), and actual driving time $\tau=\pi/4$.}
%     \label{four_1rp}
% \end{figure}
\section{Distinguishing entangled quantum states}
Here, we discuss the potential application of quantum speed limit time as a dynamical quantum witness to distinguish entangled states. We mainly focus on the Bures-angle-based measure of quantum speed limit time to show the usefulness of $\tau_{QSL}$ to distinguish quantum states. In Figs.~\ref{two_B_CB_NMAD},~\ref{three_CB} and~\ref{four_CB}, we have quantum speed limit time as a function of $\kappa \tau$ for two, three, four qubits maximally entangled states under dissipative evolution, respectively. Multi-qubit entangled states with an equal number of up and down qubits form a degenerate set of entangled states.
From Figs.~\ref{two_B_CB_NMAD},~\ref{three_CB} and~\ref{four_CB}, it is clear that maximally entangled states constructed with a different number of up and down qubits can be distinguished via  their speed limit time. These results can be extended to general $N$ qubits entangled states. \newline

As it's known, the dimension of Hilbert space for an array of $N$ qubits is $2^N$. The $2^N$ basis vectors are constructed from the different combinations of $N$ qubits in down ($\downarrow$) and up ($\uparrow$) states. Among  $2^{N}$ computational basis vectors, we have one each with all qubits  in up and down states. Similarly, there are ${}_{N}C_{r}=\begin{pmatrix}
N\\
r
\end{pmatrix}=\frac{N!}{r!(N-r)!}$ basis vectors  with $r$ qubits in downstate, and $\begin{pmatrix}
N\\
N-r\end{pmatrix}$ states with $N-r$ qubits in upstate. 
The total number of basis vectors $2^N=1+N+\begin{pmatrix}
N\\
2
\end{pmatrix}+....+\begin{pmatrix}
N\\
N-2
\end{pmatrix} +N+1$. The superposition of computational basis vectors with  an equal number of up and down states and with equal probability amplitude gives the maximally entangled GHZ basis vectors. Among the  $2^N$ vectors of entangled basis,   $2 \begin{pmatrix}
N\\
r\end{pmatrix}$ $\Bigg[2 \begin{pmatrix}
N\\
N-r\end{pmatrix}\Bigg]$ vectors form a degenerate set with  $r$ $[N-r]$ number ($r=1,..N$) of qubits with up and down state. From our work on two, three and four qubit entangled $X$ states; we show that quantum speed limit time can be availed as a dynamical witness to distinguish multi-qubit  states with different degeneracy.
\section{Conclusions}
This work investigated the role of coherence and mixedness in determining the speed limit time between the quantum states for both dephasing and dissipative processes. $\tau_{QSL}$ calculations were based on relative purity and Bures angle measures. Using maximally coherent pure, mixed and multi-qubit  $X$ states as initial states,  the role of initial coherence in their speed of evolution for both dephasing and dissipative processes was discussed. We showed that initial coherence and the trade-off between coherence and mixedness play a significant role in deciding the nature of quantum speed limit time.  The flow of quantum speed limit time revealed the role of the non-zero value of initial coherence under information backflow conditions for the dissipative process. The coherence-mixing trade-off for a single qubit state was seen to be invariant under dephasing evolution, which was not the case for the dissipative scenario. The parametric trajectory of speed limit time vividly depicts the difference in the evolution of pure and mixed initial states, and  this  was used to distinguish between the unital and non-unital channels discussed in this work. Our investigation of quantum speed limit time on multi-qubit $X$ states revealed its potential application as a dynamical witness to distinguish entangled quantum states among different degenerate sets. 
\section*{Acknowledgement}
SB acknowledges the support from the Interdisciplinary Cyber-Physical Systems (ICPS) programme of the Department of Science and Technology (DST), India, Grant No.: DST/ICPS/QuST/Theme-1/2019/6.

\bibliography{apssamp}% Produces the bibliography via BibTeX.
%\printbibliography
\section*{Appendix}
\setcounter{equation}{0}
\appendix*
%\chapter{An appendix}
%\renewcommand{\theequation}{I\Appendix.\arabic{equation}}
The details of the noisy quantum channels used in this work are in the appendix. We consider both dephasing and dissipative non-Markovian quantum noise.
\subsection{ Dephasing quantum channels }
The dynamics of the quantum system under dephasing, unital process~\cite{Shrikant2018} is given by the master equation,
\begin{equation}
    \dot{\rho_{t}}=\gamma(t)(s_{z}\rho_{t} s_{z}-\rho_{t}).
    \label{depmas}
\end{equation}
For the  initial state expressed,
\begin{equation}
\rho_{0}=\frac{1}{2}
    \begin{pmatrix}
    1+\eta_{z} & \eta_{x}-i \eta_{y}  \\
  \eta_{x}+i \eta_{y}  & 1-\eta_{z},
\end{pmatrix}
\label{intstate}
\end{equation}
for $\eta=(\eta_{x},\eta_{y},\eta_{z})$, $\eta \epsilon \mathcal{R}^3 $, and $||\eta||\leq1$, 
the time-dependent reduced  density matrix is as follows,
\begin{equation}
\rho_{t}=\frac{1}{2}
    \begin{pmatrix}
    1+\eta_{z} & (\eta_{x}-i \eta_{y}) p_{t}~  \\
  (\eta_{x}+i \eta_{y})p_{t}  & 1-\eta_{z}
\end{pmatrix}.
\label{dephfinal}
\end{equation}
The decoherence function $p_t=e^{-2\Lambda_{t}}$, $\Lambda_{t}=\int_{0}^{t}\gamma(t)dt$, where the decoherence rate $\gamma(t)=-\frac{\dot{p_t}}{2p_t}$.\newline
The measure of non-Markovianity $\mathcal{N}_{\mathcal{L}}$ is defined as a deviation from temporal self-similarity~\cite{shrikant2020},
    \begin{equation}
    \mathcal{N}_{\mathcal{L}}=min_{\mathcal{L}^{*}}\frac{1}{T} \int_{0}^{T}\vert\vert \mathcal{L}(t)-\mathcal{L}^{*}\vert\vert dt,
    \label{NMMSR}
    \end{equation}
where, $\vert\vert A\vert\vert=\textrm{tr}\sqrt{A A^\dag}$ is the trace norm of the operator,  $\mathcal{L}(t)$ and $\mathcal{L}^{*}$ are the  generators of non-Markovian and Markovian evolution, respectively. $\mathcal{N}_{\mathcal{L}}=0$ iff the channel is a quantum dynamical semigroup (QDS) and is greater than zero for a deviation from QDS. $\mathcal{L}-\mathcal{L^*}=(\gamma^*-\gamma)(\vert\phi^+\rangle\langle\phi^+\vert-\vert\phi^-\rangle\langle\phi^-\vert)$, and $\vert\phi^{\pm}\rangle$ is the Bell diagonal states.\newline
\subsubsection{CP-divisible phase damping channel; Modified Ornstein–Uhlenbeck noise (OUN)}% correction is required, this is not modified OUN
OUN noisy channel is Markovian under CP-divisibility~\cite{yu2010} criteria but is non-Markovian due to the presence of memory.
% \begin{equation}
%   p(t) =
%     \begin{cases}
%       e^{\frac{-\mu}{2}\{t+\frac{1}{\Gamma}(e^{(-\Gamma t)-1)}\}}& \text{OUN}\\~\\
%       e^{-\frac{\mu t(\Gamma t+2)}{2(\Gamma t+1)^2}} & \text{PLN}.\\
%     \end{cases}       
% \end{equation}
The decoherence function of OUN is,
\begin{equation}
    p_t=  e^{\frac{-\kappa}{2}\{t+\frac{1}{\lambda}(e^{-\lambda t}-1)\}},
    \label{OUNfn}
\end{equation}
where $\lambda^{-1}\approx\tau_{r}$ defines the reservoir's finite correlation time, and $\kappa$ is the coupling strength related to the qubit's relaxation time.
The decoherence rate is as follows,
% \begin{equation}
%   \gamma(t) =
%     \begin{cases}
%       \frac{\mu (1-e^{-\Gamma t})}{4}& \text{OUN}\\~\\
%       \frac{\mu}{2(\Gamma t+1)^3} & \text{PLN}\\
%     \end{cases},      
% \end{equation}
\begin{equation}
    \gamma(t) =
          \frac{\kappa (1-e^{-\lambda t})}{4},
\end{equation}
this  channel is not CP-indivisible but non-Markovian according to the measure in Eq.~\ref{NMMSR}, and $\gamma(t)$ is positive for all values of $t$. Due to the  system-environment interaction in the cases of OUN, the Markovian regime is achieved in the limit $\frac{1}{\lambda}\rightarrow\infty$, the corresponding decoherence function for OUN  channel is $p^{*}(t)=e^{-\kappa t/2}$.
\subsubsection{P-indivisible phase damping; Random Telegraph noise (RTN)}
RTN channel~\cite{mazzola2011,Pradeep}, is non-Markovian according to information backflow and CP-divisibility criteria. The decoherence function in this case has the form $p_t=e^{-\kappa t}\Big[\cos\Bigg(\sqrt{[(\frac{2c}{\kappa})^2-1]}\kappa t\Bigg)+\frac{\sin\Bigg(\sqrt{[(\frac{2c}{\kappa})^2-1]}\kappa t\Bigg)}{\sqrt{(\frac{2c}{\kappa})^2-1}}\Big]$.
The frequency of the harmonic oscillators is $\sqrt{(\frac{2c}{\kappa})^2-1}$. The parameters $c$ and $\kappa$ correspond to the strength of the system-environment coupling and the fluctuation rate of the RTN, respectively. The noise parameter describes two regimes of systems dynamics, for $\frac{c}{\kappa}<0.5$, the channel corresponds to the Markovian dynamics, the purely damping regime and damped oscillations for $\frac{c}{\kappa}>0.5$ (damped oscillations) corresponds the non-Markovian evolution.
\subsection{Non-Markovian amplitude damping channel (NMAD)} The solvable Jaynes-Cummings model for a two-level system resonantly coupled to a leaky single-mode cavity is considered. The reduced system's dynamics are given as
\begin{equation}
\dot{\rho}=\gamma_{i}\Bigg(\sigma_{-}\rho_{t}\sigma_{+}-\frac{1}{2}\sigma_{+}\sigma_{-}\rho_{t}-\frac{1}{2}\rho_{t}\sigma_{+}\sigma_{-}\Bigg),
\end{equation}
where $\sigma_{\pm}=\frac{1}{2}(\sigma_{x}\mp i\sigma_{y})$, and exhibits a non-unital evolution.\newline
For initial state Eq.~\ref{intstate}, time-dependent reduced density matrix is given,
\begin{equation}
\rho_{t}=\frac{1}{2}
    \begin{pmatrix}
    2-(1-\eta_{z})\vert p_{t}\vert^2 & (\eta_{x}-i \eta_{y})p_{t} \\
  (\eta_{x}+i \eta_{y})p_{t}  & (1-\eta_{z})\vert p_{t}\vert^2
\end{pmatrix},
\label{adfinal}
\end{equation}
where $p_{t}=e^{-\Lambda_{t}/2}$, $\Lambda_{t}=\int_{0}^{t}\gamma(t)dt$. We have, $\gamma(t)=-\frac{\dot{2p_{t}}}{p_t}$, and $p_t=e^{-\lambda t/2}\big(\sinh(dt/2)d \cosh(dt/2)+\Gamma\sinh(dt/2)\big)$ with time-dependent decoherence rate,
\begin{equation}
    \gamma(t)=\frac{2\kappa\lambda \sinh(dt/2)}{d \cosh(dt/2)+\lambda\sinh(dt/2)},
\end{equation}
where $d=\sqrt{\lambda^2-2\kappa\lambda}$, $\lambda$ is the spectral width of the reservoir, and $\kappa$ is the coupling strength between the qubit and the cavity field.\newline
Measure of non-Markovianity $\mathcal{N}_{\mathcal{L}}=\min_{\gamma^{*}}\frac{1}{\tau}\int_{0}^{\tau}\vert\gamma(t)-\gamma^{*}\vert(1+\sqrt{2})dt$. 
\end{document}